\documentclass[10pt, conference]{IEEEtran}
\usepackage{cite}
\usepackage{amsmath,amssymb,amsfonts}
\usepackage{textcomp}
\def\BibTeX{{\rm B\kern-.05em{\sc i\kern-.025em b}\kern-.08em
    T\kern-.1667em\lower.7ex\hbox{E}\kern-.125emX}}
\usepackage[usenames,dvipsnames]{xcolor}
\usepackage{graphicx}
\usepackage{float}

\usepackage{braket}
\usepackage{booktabs}
\usepackage{physics}
\usepackage{bm}
\usepackage{tabularx}

\usepackage[utf8]{inputenc}
\usepackage[linesnumbered,ruled,vlined]{algorithm2e}
\SetKwInput{KwInput}{Input}
\SetKwInput{KwOutput}{Output}
\SetFuncSty{textsc}
\SetKwFunction{SelectSubgraph}{SelectSubgraph}
\SetKwFunction{SolveSubgraph}{SolveSubgraph}

\usepackage[bookmarks=true,colorlinks,citecolor=blue,urlcolor=blue]{hyperref}

\begin{document}

\author{\IEEEauthorblockN{1\textsuperscript{st} Ji\v r\' i Guth Jarkovsk\' y}
\IEEEauthorblockA{\textit{IQM Quantum Computers}\\
München, Germany \\
jiri.guthjarkovsky@iqm.tech}
\and
\IEEEauthorblockN{2\textsuperscript{nd} Patricia Bickert}
\IEEEauthorblockA{\textit{Deutsche Bahn AG / DB Systel}\\
Frankfurt am Main, Germany \\
patricia.bickert@deutschebahn.com}
\and
\IEEEauthorblockN{3\textsuperscript{rd} Elisabeth Wybo}
\IEEEauthorblockA{\textit{IQM Quantum Computers}\\
München, Germany \\
elisabeth.wybo@iqm.tech}
\and
\IEEEauthorblockN{4\textsuperscript{th} Martin Leib}
\IEEEauthorblockA{\textit{IQM Quantum Computers}\\
München, Germany \\
martin.leib@iqm.tech}
}

\title{Rolling Stock Planning Using the Quantum Approximate Optimization Algorithm}

\maketitle

\begin{abstract}
Rolling stock planning is a complex optimization problem in railway management that involves assigning physical trains to scheduled trips while minimizing operational costs. In this work, we address a specific instance of this problem featuring 190 trips over two days, subject to constraints such as mandatory maintenance stops. We reformulate the problem as a Maximum-Weight Independent Set (MWIS) problem on a graph where nodes represent feasible train cycles. To handle the computational complexity of the large search space, we propose a hybrid divide-and-conquer algorithm. This approach iteratively selects subgraphs and solves the MWIS problem using various solvers, including exact classical methods and the Quantum Approximate Optimization Algorithm (QAOA). We evaluate the algorithm's performance by comparing these methods and analyzing the scaling with respect to subgraph size, with QAOA assessed through both classical simulation and execution on a quantum device (IQM Emerald). Our results indicate that increasing the subgraph size generally improves solution quality, demonstrating that the hybrid framework can effectively bridge the gap between polynomial-time approximate solvers and exponential-time exact methods.
\end{abstract}


\section{Introduction}

\subsection{Railway-related optimization problems}

In the broad field of railway management and operations, there are many related, difficult optimization problems. These problems arise from the need to coordinate shared infrastructure, heterogeneous resources, safety constraints, and stochastic disturbances, often in real time.

The first problem, chronologically, is timetabling. This refers to building a timetable for the trains, while responding to customer demand and optimizing for factors such as route / rolling stock utilization and robustness against disruptions.

Rolling stock planning and fleet assignment problems then handle the assignment of physical trains (rolling stock) to service the trips in the timetable. This problem brings new constraints, such as regular maintenance of rolling stock, and additional operating costs, such as energy use. This paper tackles a specific case of this optimization problem.

In the case of unforeseen circumstances (e.g., a train breakdown), the trains need to be rerouted on a short timescale. This problem then becomes much more about the exact topology of the railroad network and the cost function also typically changes from saved fuel to something like a cumulative delay (as disturbances in a tightly-optimized railroad network tend to cascade).

There are naturally many other categories of optimization problems in a big industrial railway company such as Deutsche Bahn (DB). DB investigates the optimization of its assets like rails, trains, and employee scheduling with hybrid systems (HPC/QC) for different use cases, but a detailed exploration of other categories than trains is beyond the scope of this work.

\subsection{Related work}
Building on the railway optimization problems outlined above, several works have explored quantum approaches to these tasks by formulating them as QUBO or HUBO problems suitable for quantum annealers or gate-based algorithms such as QAOA. In the following, we review the most relevant contributions and position our work within this emerging literature.

In chapter 4 of \cite{Grange2024}, Grange tackles the timetabling problem by simplifying it to the Set Cover Problem (SCP) and the Extended Bin Packing Problem, providing the QUBO / HUBO formulation of both and some simulated results.

For the present paper, the work of Bickert et al. \cite{Bickert2023} is particularly relevant for us as their work tackles a problem very similar to the one presented here. The difference is that in \cite{Bickert2023} the authors use constrained programming (CP) and quantum annealing to directly solve the problem, whereas here we devise a hybrid algorithm using QAOA.

In a series of works \cite{Domino2022, Domino2023}, Domino et al. examine rail rescheduling caused by delays on a small rail network, formulating the problem as a quadratic (QUBO) and higher-order (HUBO) unconstrained binary optimization cost function and solving it with a quantum annealer.

We would also like to bring attention to the works of Vikstal et al. \cite{Vikstal2020} and Gili et al. \cite{Gili2024}. These works tackle the Tail Assignment Problem (TAP). This problem concerns the assignment of tails (physical aircraft) to scheduled flight routes. Mathematically, it is very similar to the rolling stock planning problem we tackle in this work.

The typical size of these problems means that they often have to be heavily simplified in order to be able to either run an experiment or classically simulate the results of a quantum device. With the hybrid divide-and-conquer algorithm proposed in this paper, we show a way forward even before quantum solvers reach the size to solve the full problem directly.

Beyond railway applications, there has also been recent interest in applying and analyzing QAOA for maximum independent set problems. Wybo and Leib \cite{Wybo2025} study the approximation performance of QAOA on random graph ensembles, while Wybo et al. \cite{Wybo2026} propose a scalable hybrid quantum-classical approach that combines QAOA-derived information with a greedy MIS construction. These works focus on generic MIS instances rather than railway optimization problems, but share the goal of extending the applicability of QAOA beyond problem sizes directly accessible to current quantum hardware.

\subsection{Paper Outline}
The remainder of this paper is structured as follows. In Sec.~\ref{sec:problem_formulation}, we introduce the rolling stock planning problem. Sec.~\ref{sec:hybrid} presents our hybrid solution approach based on a mapping to the maximum-weight independent set problem and a divide-and-conquer algorithm. Sec.~\ref{sec:results} reports numerical results comparing different subgraph solvers and analyzing the impact of subgraph size. Sec.~\ref{sec:discussion} discusses the algorithm’s components and performance, and Sec.~\ref{sec:outlook} concludes with an outlook.

\section{Problem Description} \label{sec:problem_formulation}
Rolling stock planning is a constrained optimization problem. The core of the problem is a timetable containing $N$ \emph{trips} (covering $N_d$ days), all of which need to be \emph{serviced}. Each trip is defined by a departure station, an arrival station, and a scheduled departure time. Trips begin in one of $N_c$ stations and end in a different one. We are given information about the railroad network, i.e., the distances between any two stations (both the distance in kilometers and the time it takes to travel by train). A definition of both the timetable and the railroad network fully describe an instance of the rolling stock planning problem class.

For a solution of the problem, all trips of the timetable have to be serviced by trains traveling through the railroad network. Each train travels in a sequence of trips. It has to end its $N_d$-day journey at the same place where it started, so that the solution can be repeated over longer time scales with an $N_d$-day period. We define the sequence of trips that a train undertakes a \emph{cycle}. Every cycle must include a stop in a particular station (the \emph{maintenance station}) during which the train undergoes a mandatory maintenance break for some time $t$. Additionally, no cycle can be longer than $l_{\text{max}}$, the distance limit before maintenance is required.

All cycles need to fulfill some common-sense constraints; a train can only start a trip after it has finished the previous one, and it can only start from the location where the previous trip ended. The cycles selected as part of the solution, taken together across all trains, must service all trips in the timetable. To make that possible, the trains are allowed to take \emph{empty trips}. An empty trip is a trip not in the timetable (and therefore not transporting any passengers). It can be taken between any two stations at any time. Despite this freedom, we may assume without loss of generality that empty trips occur only immediately after a scheduled trip or a maintenance break. This follows from the following three cases:
\begin{enumerate}
    \item If an empty trip was going to take place at the beginning of a cycle, then the cycle may instead be started in the destination of the initial empty trip and an extra empty trip may be added at the end of the cycle (to make it a cycle).
    \item If an empty trip was going to take place some time after finishing the preceding trip, the solution quality is unaffected by moving the empty trip back in time to immediately follow the preceding trip, which is possible since we are not accounting for the capacity constraints of the railroad network. 
    \item If an empty trip follows another empty trip, we can cancel both of these trips and replace them with a new empty trip from the origin of the first empty trip directly to the destination of the second empty trip.
\end{enumerate}

Empty trips ensure that feasible solutions exist. However, they incur a cost in the form of \emph{empty kilometers}, i.e., distance traveled by trains without transporting passengers. These empty kilometers can be regarded as a proxy for operational costs such as energy consumption and personnel costs.

Therein lies the optimization problem. The goal is to design the cycles in such a way that all timetabled trips are serviced, all constraints are satisfied, and the number of empty kilometers is minimized.

We note that in this work we do not account for capacity constraints of the railroad network, i.e., the number of trains that can be ``parked'' in each station at the same time is unlimited and so is the number of trains that can travel along a given route at any time.

\section{Hybrid Algorithm Description} \label{sec:hybrid}

We devise a hybrid algorithm that applies a quantum solver to subproblems within a larger classical iterative framework. The algorithm we describe in the following might not find a solution that services all trips of the timetable. In this case we iteratively remove the already serviced trips from the timetable in a classical outer loop calling the algorithm again, until all trips in the timetable are serviced.

\subsection{Mapping to a Maximum-Weight Independent Set Problem}
\label{subsec:algo_reform}

At the beginning of the algorithm we map the problem instance to a \emph{maximum-weight independent set} (MWIS) problem. To this end, we generate cycles that a train could take, given the existing timetable. Cycles are generated with the goal that the algorithm is capable of finding solutions to a maximal subset of the trips in the timetable. We represent each cycle by a node in a graph, connecting nodes (cycles) with an edge if they are mutually incompatible (i.e., two cycles servicing the same scheduled trip). The nodes are assigned weights that reflect the cost which we are trying to optimize for. Now, the solution to the optimization problem is encoded in an MWIS problem. That is, an \emph{independent set} (set of nodes not directly connected by any edges) whose total node weight is maximal. We may formalize the MWIS problem as a constrained optimization problem in the following form, 
\begin{align}
\max_{x \in \{0,1\}^{|V|}} \quad & C(x) = \sum_{i \in V} w_i x_i \\
\text{subject to} \quad & \sum_{(i,j) \in E} x_i x_j = 0.
\end{align}
Here $V$ and $E$ are the sets of vertices and edges of the graph. The weight of vertex $i$ is $w_i$ and $x_i \in \{0,1\}$ is a binary variable indicating whether vertex $i$ is included in the selected set.

\paragraph{Generating Cycles}

There are various ways to generate suitable cycles. In this work, we use the following function, operating in two modes, called mode 1 (no empty trips) and mode 2 (empty trips allowed):

\begin{enumerate}
    \item Begin in a station at the beginning of the first day.
    \item Branch over all feasible subsequent actions from the current station at the current time:
    \begin{itemize}
        \item Branch on all available scheduled trips departing from the current station after the current time.
        \item If operating in mode 2, also branch on all available empty trips (respecting the restrictions defined in Section \ref{sec:problem_formulation}). If operating in mode 1, do not.
        \item If the current station is the \textbf{maintenance location} and maintenance has not yet been performed, also branch on performing maintenance
    \end{itemize}
    \item Update the current time and location based on the selected action.
    \item Repeat steps 2-3 until there are no more actions to take (it is near the end of the scheduling period). If the final location is the same as the initial location, the maintenance has been performed and the cycle does not exceed the \textbf{maintenance distance}, save it as a valid cycle. Otherwise discard it.
    \item Continue until all branches have been explored (either discarded or saved as valid cycles).
    \item Repeat steps 1-5 for all possible starting stations.
\end{enumerate}

The data set contains $N$ scheduled trips over $N_d$ days. To control the computational complexity of cycle enumeration, we switch the mode used for generating cycles mid-algorithm. At first, we generate cycles using the above process in mode 1 (no empty trips). When the number of unserviced trips falls below a threshold $T$ (40 in our case), or when no further cycles can be formed from the remaining trips (whichever comes first), we start generating cycles in mode 2 (empty trips allowed). Recall that each scheduled trip can be followed by one of $N_c - 1$ different empty trips. Thus, allowing empty trips effectively multiplies the possible pool of trips for cycle generation by $N_c$.

By appropriately defining the weights of the MWIS problem, $w_i$, we are capable of defining the optimization target of the rolling stock planning problem. For the purpose of this work we have chosen a target of the optimization problem that rewards passenger coverage (full kilometers) while penalizing operational inefficiency (empty kilometers). Specifically, $w_i=2d_{\text{full}} - d_{\text{empty}}$, where $d_{\text{full}}$ is the distance traveled with passengers and $d_{\text{empty}}$ is the empty distance. The coefficient 2 in front of $d_{\text{full}}$ was chosen heuristically to prioritize the servicing of scheduled trips over pure cost reduction.

As mentioned above, because the cycle generation step explores only a restricted subset of all feasible cycles (in particular, by delaying the introduction of empty trips), and the MWIS solver may return suboptimal solutions, a single run of the algorithm may fail to service all trips in the timetable. To address this, we adopt an iterative approach: after each run, the trips that have already been serviced are removed, and a new problem instance is constructed on the remaining trips. Repeating this process ensures that all trips are eventually covered.

Although the MWIS formulation provides a convenient representation of the problem, it introduces a lot of decision variables (one per cycle). However, the number of all possible cycles only scales polynomially with the size of the timetable, although with a high degree, see Appendix~\ref{sec:n_cycles}. Even for moderately sized time-tables (with approx. 200 trips), the resulting MWIS graph is too big for any generic MWIS solver (see Appendix \ref{sec:n_cycles}). So, instead of solving the problem all at once, we design a divide-and-conquer algorithm to approach the problem.

\subsection{Quantum Divide-and-Conquer}
\label{subsec:solving_mwis}

Our main contribution is a divide-and-conquer algorithm with a quantum subroutine for approximately solving the large MWIS problem arising from rolling stock planning.

This design allows quantum algorithms to be applied to subproblems that are compatible with current hardware limitations, while a classical outer loop orchestrates the overall optimization. The algorithm is described in Algorithm~\ref{alg:DnC}.


\begin{algorithm}
\caption{Divide-and-Conquer Algorithm}
\label{alg:DnC}
\KwIn{Graph $G=(V,E)$ with vertex weights $w: V \rightarrow \mathbb{R}^+$}
\KwOut{Independent set $I \subseteq V$ with large total weight}

$I \gets \emptyset$ \;
$G' \gets G$ \;

\While{$G'$ is not empty}{
    \tcp{Select a subgraph of size $k$}
    $G'' \gets \SelectSubgraph{$G', k$}$ \;
    
    $I^\prime \gets \SolveSubgraph{$G''$}$ \;
    
    $I \gets I \cup I'$ \;
    
    \tcp{Remove the set $I'$ and all its neighbors from the graph}
    $G' \gets G' \setminus \bigl(I' \cup N_{G'}(I')\bigr)$ \;
}

\Return $I$

\end{algorithm}

\subsubsection{Greedy Algorithm}
\label{sec:greedy_algo}

As an illustrative, extreme case of the above-described divide-and-conquer algorithm we consider the \emph{greedy algorithm} (see e.g., \cite{Sakai2003}). 

The greedy algorithm iteratively selects a subgraph of size 1, i.e. a node, with the highest weight-to-node-degree ratio and adds it to the independent set. This ratio balances the node weight against its connectivity, since selecting a node also excludes all of its neighbors from being chosen. Finally, the MWIS graph is updated as described in~\ref{alg:DnC}.

For the corresponding pseudocode, see Algorithm~\ref{alg:greedy}.

\begin{algorithm}
\caption{Greedy Maximum-Weight Independent Set}
\label{alg:greedy}
\KwIn{Graph $G=(V,E)$ with vertex weights $w: V \rightarrow \mathbb{R}^+$}
\KwOut{Independent set $I \subseteq V$ with large total weight}

$I \gets \emptyset$ \;
$G' \gets G$ \;

\While{$G'$ is not empty}{
    \tcp{Select vertex $v$ with highest weight-to-degree ratio}
    $v \gets \arg\max_{u \in V(G')} \frac{w(u)}{deg_{G'}(u) + 1}$ \;
    
    $I \gets I \cup \{v\}$ \;
    
    \tcp{Remove $v$ and all its neighbors from the graph}
    $G' \gets G' \setminus \bigl(\{v\} \cup N_{G'}(v)\bigr)$ \;
}

\Return $I$
\end{algorithm}

The greedy algorithm serves as a classical baseline. We describe it here for reference, and later we will compare it against our approach.

\subsubsection{Divide and Conquer in Detail}
\label{sec:algo_in_detail}

\paragraph{Subgraph Selection}
A key design choice in the proposed hybrid framework is the method used to select subgraphs of the global MWIS instance, the function \textsc{SelectSubgraph} from pseudocode~\ref{alg:DnC}. This choice directly influences the balance between exploration of the solution space and exploitation of high-quality candidate cycles.

A natural baseline strategy is to select subgraphs uniformly at random from the set of all cycle nodes. While this approach is unbiased, it tends to yield subgraphs dominated by low-quality or suboptimal cycles. Consequently, the MWIS solver operating on such subgraphs is more likely to include these inferior candidates in the partial solution, which negatively impacts the overall solution quality.

At the opposite extreme, one may deterministically select nodes with the largest weights. This biases the subgraph toward high-quality cycles and typically improves local solution quality. However, this strategy effectively reduces the method to a variant of the greedy approach. Indeed, when the subgraph consists predominantly of the highest-weight nodes, the MWIS solution on the subgraph often coincides with the choices made by the greedy algorithm, thereby limiting any potential advantage of solving larger subproblems.

These observations motivate the need for an intermediate selection strategy that preferentially includes promising cycles while still promoting diversity in the explored solution space. In this work, we adopt a heuristic based on the number of serviced trips per cycle. Specifically, subgraphs are constructed from cycles that maximize the number of full (passenger-carrying) trips. This criterion is correlated with, but not identical to, the node weights used in the MWIS formulation. As a result, it tends to select cycles that are near-optimal with respect to the objective function, while avoiding an excessive concentration on the highest-weight nodes alone. Empirically, this selection strategy yields the best overall performance among the methods considered.

Once a subset of nodes has been selected, we solve the MWIS on it. This allows us to insert a quantum MWIS solver into the overall workflow.

\paragraph{QAOA MWIS Solver}
We use the quantum approximate optimization algorithm (QAOA) \cite{farhiQuantumApproximateOptimization2014} as an MWIS solver.
This step refers to the function \textsc{SolveSubgraph} from pseudocode~\ref{alg:DnC}, the function to solve the MWIS problem on the selected subgraph.

QAOA is an variational quantum algorithm that starts with mapping the specific optimization problem to a diagonal spin glass Hamiltonian. 

\begin{align}
C(x) \rightarrow H 
=
-\sum_{i \in V} \frac{w_i}{2} (z_i + 1)
+
\lambda \sum_{(i,j)\in E} (z_i + 1) (z_j + 1)\,,
\end{align}
where $\lambda > 0$ is the penalty applied for each edge whose endpoints are both selected and $z_i$ is the Pauli-z operator acting on qubit $i$. To ensure that any optimal solution of the diagonal spin glass Hamiltonian corresponds to a valid independent set, the penalty parameter must satisfy $\lambda \ge \max_i 4 w_i$. In our case, we set $\lambda = \max_i 4 w_i$.

With this in mind any variational ansatz where we prepare the parameter vectors $\bm{\beta}=(\beta_1, \dots, \beta_p)$ and $\bm{\gamma}=(\gamma_1, \dots, \gamma_p)$ according to 
\begin{align}
{\bm{\beta},\bm{\gamma}} = {\operatorname*{argmin}_{\bm{\beta'},  \bm{\gamma'} }} \left\langle \bm{\beta'},  \bm{\gamma'} |H|\bm{\beta'}, \bm{\gamma'} \right \rangle\,,
\end{align}
would provide low energy solutions to the diagonal spin-glass Hamiltonian and therefore to our MWIS problem. To get these one would have to prepare the variational state on the quantum computer and measure in the computational basis. The QAOA Ansatz does additionally include information of the optimization problem in the following way,
\begin{align}
|\bm{\beta}, \bm{\gamma}\rangle = \prod\limits_{i = 1}^p \exp(-\mathrm{i}\beta_i H_{\text{driver}}) \exp(-\mathrm{i}\gamma_i H) |+\rangle\,,
\end{align}
with the problem agnostic driver Hamiltonian $H_{\text{driver}} = \sum_i x_i$, where $x_i$ is the Pauli-x operator acting on qubit $i$. 
It is a Trotterized version of the quantum annealing algorithm \cite{kadowakiQuantumAnnealingTransverse1998}, where the Trotter-step lengths are variationally defined. In this way one can show that if the number of Trotter steps is infinitely large $p\rightarrow \infty$ the gate based algorithm is approximating quantum annealing and does prepare the ground state of the diagonal spin-glass Hamiltonian exactly and therefore finds the optimal solution to the MWIS problem. Furthermore it can be shown, with techniques from optimal control theory, that the specific Trotterised setting of the QAOA algorithm is the optimal solution given finite time \cite{yangOptimizingVariationalQuantum2017}.

\paragraph{Pruning}

The results from QAOA (either simulated or actually run on a quantum computer) and random guessing are generally not valid independent sets. To create proper independent sets from them, we employ a \emph{pruning} procedure. Here, pruning refers to an efficient algorithm for removing nodes from a non-independent set, in order to make it independent. Pruning helps to make an infeasible ``solution candidate'' feasible, but it cannot be used to solve the optimization problem (by starting from the set of all nodes), since solving MWIS is NP-hard.

\begin{enumerate}
    \item Construct the subgraph induced by the sampled solution. If this graph has 0 edges, the solution is already an independent set and therefore no pruning is required. In that case skip to step 5.
    \item To each node with edges, assign a ratio of the node's weight divided by its degree.
    \item Remove the node with the lowest ratio and all its edges.
    \item Repeat steps 2-3 as long as there are edges in the graph.
    \item Return the nodes of the graph as a pruned solution.
\end{enumerate}

\paragraph{Global Graph Update}
Once a feasible solution for the subgraph is obtained, we update the global graph to reflect the serviced trips. All nodes corresponding to the cycles selected in the subgraph solution are removed from the global graph. Additionally, all nodes sharing an edge with the selected nodes (i.e., servicing any trip already serviced) are removed to ensure conflict constraints are met. The algorithm then iterates, selecting a new subgraph from the remaining nodes until all nodes are either in the selected independent set or discarded.

\section{Results} \label{sec:results}

We empirically evaluate the proposed algorithm under different subgraph solver choices and subgraph sizes. We first describe the experimental setup, and then the results.

\subsection{Experimental Setup}

\subsubsection{Problem Parameters}

The parameters of our particular problem instance are summarized in Table \ref{tab:problem_parameters}.

\begin{table}[t]
\centering
\caption{Problem instance parameters}
\label{tab:problem_parameters}
\begin{tabularx}{\linewidth}{lXX}
\toprule
Parameter & Description & Value \\
\midrule
$N_c$ & Number of stations & 5  \\
- & Station set & {Cologne, Munich, Berlin, Frankfurt, Hamburg} \\
- & Maintenance station & Hamburg \\
 $N_d$ & Number of days & 2 \\
 $N$ & Number of trips in the timetable & 190 \\
 $t$ & Maintenance duration & 2 hours \\
 $l_{\text{max}}$ & Maximum length of cycle & 4000 km \\
 
\bottomrule
\end{tabularx}
\end{table}

\subsection{Comparison of Subgraph Solvers}
\label{sec:comparison_of_solvers}

\begin{figure*}[h]
    \centering
    \includegraphics{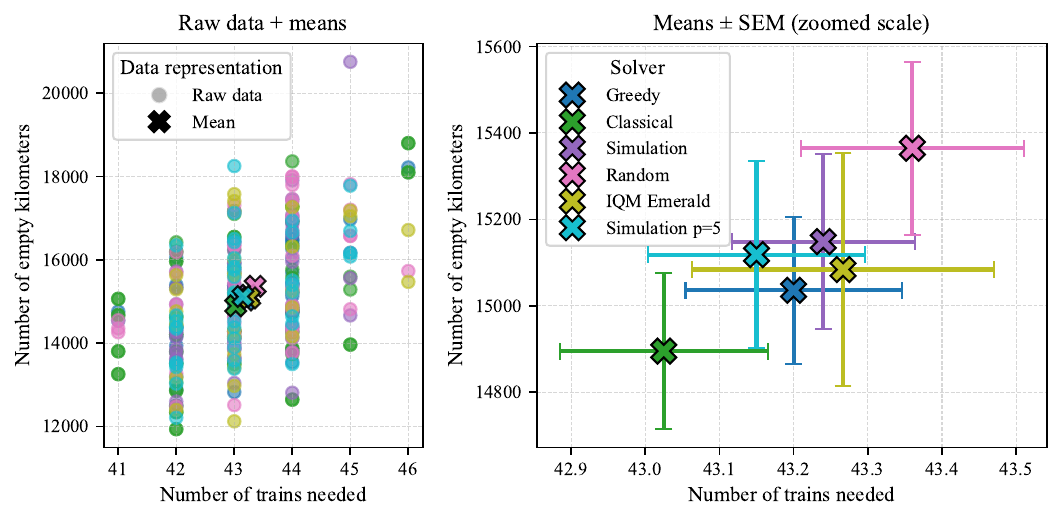}
    \caption{Results showing the performance of various methods to solve the subgraph MWIS, with subgraph size kept fixed at 20 (and the greedy algorithm \ref{sec:greedy_algo} as a benchmark). The left panel shows the data points from each individual full algorithm run, while the right panel shows the average performance of each method with standard error of the mean (SEM). \emph{IQM Emerald} refers to results obtained from QAOA runs on IQM Emerald quantum computer. For more details about the solvers, see \ref{sec:comparison_of_solvers}.}
    \label{fig:compare_methods}
\end{figure*}

First, we observe how the algorithm performs when using different methods to solve the sub-problems. In this subsection, we fix the subgraph size to 20 and compare the performance of an exact classical MWIS solver~\footnote{In particular, the function \texttt{max\_weight\_clique} from NetworkX~\cite{networkx}, applied to the complement graph.}, a simulation of $p=1$ and $p=5$ QAOA with pruning, a quantum hardware run of $p=1$ QAOA with pruning and random sampling with pruning. The $p=1$ QAOA is trained using analytical formulas for QAOA expectation values \cite{Ozaeta2022} whereas the $p=5$ QAOA uses linear-ramp parameters \cite{Montañez-Barrera2025}.  The exact solver yields only one solution, but for the sampling solvers (QAOA and random guessing), we take 10,000 samples, prune them, and pick the best one. Fig.~\ref{fig:compare_methods} illustrates the resulting solution quality across these methods, also compared to a simple greedy algorithm.

In Fig.~\ref{fig:compare_methods}, we plot the number of empty kilometers against the number of trains required (equal to the number of selected cycles). Although fleet size is not part of the optimization objective, it is a relevant operational metric, as fewer trains correspond to lower resource usage. This representation allows us to compare the methods in terms of both cost (empty kilometers) and resource consumption (fleet size).

For each method, we solve the entire problem multiple times (with varying seed to decide ties within the algorithm). Each of those runs is plotted in the figure (as a circle). The average of all data points using each method is marked with the cross.

The first observation is that there is a large variance in the quality of the results. Due to the structure of the algorithms, a suboptimal choice of cycles early in the calculation could possibly lead to the removal of some high-quality cycles from the graph and then to a poor overall solution quality.

The second observation is that the average performance of the different methods is very similar. For such small subgraph sizes, the performance difference between our hybrid algorithm and the greedy algorithm appears negligible. There seems to be a small improvement of the overall solution quality as a function of the subgraph solver quality, with random sampling on one end and exact solver on the other end. We expect the quality of the subgraph solver to have bigger impact as the subgraph size is increased.

Despite the very small improvement between different subgraph solvers, we do see the tendency of higher depth QAOA $p> 1$ to approach the solution quality of the exact solver. This is remarkable because QAOA for finite $p$ is a polynomial runtime algorithm compared to the exact solver which is an exponential runtime algorithm. This suggests a quantum advantage scenario where the depth of the QAOA algorithm $p$ would only grow polynomially with the subgraph size for an approximately optimal solution. In the following section we highlight how this approximately optimal subgraph solution would yield better overall solutions with increasing subgraph size.

\subsection{Impact of Subgraph Size}
\begin{figure*}[h]
    \centering
    \includegraphics{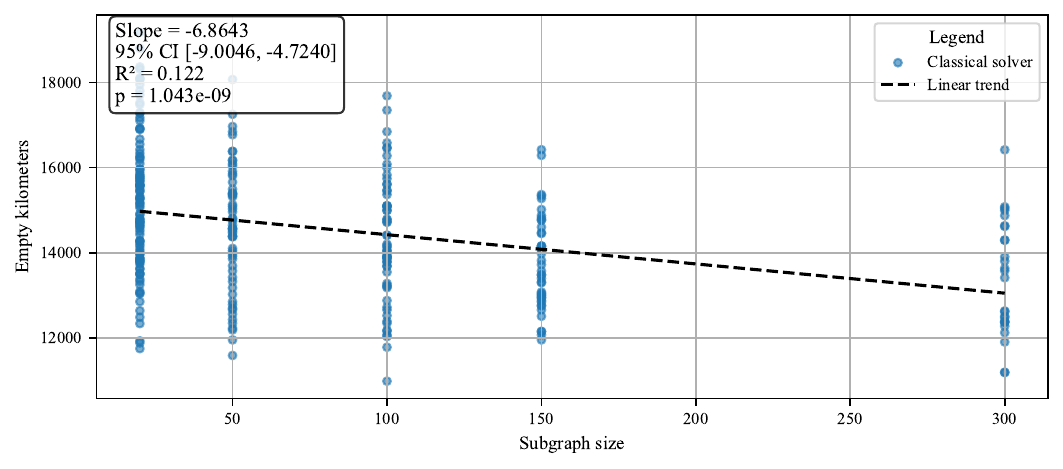}
    \caption{Results showing the performance (measured by the number of empty kilometers) of the algorithm using the classical exact MWIS solver, as a function of the subgraph size. Linear fit of the data shows a clear trend of improving solution quality with increasing subgraph size.}
    \label{fig:subset_size_trend}
\end{figure*}

Here we examine the performance of the algorithm with regard to the size of the subgraph that we select in each step, see Fig. \ref{fig:subset_size_trend}. We use exclusively a classical solver for these results, to make sure the results reflect the impact of subgraph size and not the quality of the subgraph solver itself.

As before, we see a large spread of the individual results (again seeded differently for tie breaks).

This time we see a clear trend that using a larger subgraph leads to better solutions. Fitting a linear dependency between the subgraph size and the number of empty kilometers driven, we get a negative slope with $p$-value of $1.0 \times 10^{-9}$ (see Fig. \ref{fig:subset_size_trend}), albeit the coefficient of determination $R^2$ only $0.122$ because of the large variance of solution quality on different runs of the algorithm.

\section{Discussion}\label{sec:discussion}

Comparing an exact solver with the greedy algorithm, we get two approaches at the extreme ends of a spectrum.

\begin{table}[h]
\centering
\caption{Greedy MWIS vs Exact Solver}
\label{tab:mwis_comparison}
\begin{tabular}{|l|c|c|}
\hline
\textbf{Property} & \textbf{Greedy Algorithm} & \textbf{B \& B} \\
\hline
Solution & Approximate & Exact \\
\hline
Time Complexity & $\mathcal{O}(|V|^2)$ & $\mathcal{O}(2^{|V|})$ \\
\hline
Memory & $\mathcal{O}(|V| + |E|)$ & $\mathcal{O}(2^{|V|})$ \\
\hline
\end{tabular}
\end{table}

An important feature of the greedy algorithm is that it builds the independent set iteratively based only on locally optimal decisions. In other words, it lacks the ``foresight'' to evaluate the effect of its decisions on the quality of the global solution.

Our solver attempts to bridge the gap between the fast/approximate/greedy approach and the slow/exact approach. Adapting the greedy approach, we construct the solution by picking locally-good parts. But unlike the greedy algorithm, we look at larger parts of the graph to choose which nodes to include in our solution and we utilize QAOA, to make that decision. Similar ideas have recently been explored for the maximum independent set problem \cite{Wybo2026}. Our approach differs in that it targets rolling stock planning problem instances specifically and we incorporate the classical greedy approach differently.

\subsection{Exact Solvability}

For the specific structure and size of our problem instances that we considered, an exact MWIS solver can still be used. This comes from the fact that each trip effectively represents only one constraint (``This trip can be serviced by only one cycle in the solution''), despite this corresponding to many edges in the MWIS graph. An integer-programming algorithm is then mainly bottlenecked by the number of constraints (equal to the number of trips) and not by the number of nodes in the graph (i.e., cycles). As we show in Appendix \ref{sec:n_cycles}, the number of cycles (nodes) scales polynomially with the number of trips (constraints), so technically the classical algorithm is still exponential in the number of cycles. Therefore, we expect the potential for quantum advantage for the industry size instances of the specific type of rolling stock planning problem that we considered in this article.

\subsection{Discussion of Algorithm Components and Hyperparameters}
\label{sec:discussion_parameters}

The overall performance of our hybrid algorithm is governed by a combination of fixed structural components and adjustable hyperparameters. To properly assess the potential quantum advantage of our approach, it is necessary to analyze the contribution of these individual elements to the final solution quality.

To understand algorithm performance, we need to look at the pruning procedure as well as the choice of several adjustable hyperparameters.

A key component in our pipeline, particularly for stochastic solvers such as QAOA and random guessing, is the \emph{pruning} procedure described in Section~\ref{sec:algo_in_detail}. Pruning acts as a post-processing step that converts infeasible samples into valid independent sets. In effect, pruning attempts to find a high-weight independent subset within the vertices selected by the solver, which can be viewed as approximately solving MWIS on the induced subgraph of those vertices. However, MWIS on general graphs is Poly-APX-hard \cite{Bazgan2005}, meaning that no polynomial-time algorithm can guarantee a constant-factor approximation in general. Consequently, as the subgraph size increases and the induced conflict structure becomes more complex, we expect the effectiveness of the pruning step to degrade. In other words, pruning may account for a significant portion of the observed performance for small instances, but it is unlikely to provide the same benefit at larger subgraph sizes.

The first critical hyperparameter is the threshold for introducing empty trips (fixed at $T=40$ in this study). Selecting an appropriate value is a balancing act: if the threshold is too low, the algorithm may stall or be forced to select a very unfavorable cycle. If it is higher, the solution space offered to the algorithm is wider and so one might expect better performance. However, if empty trips are allowed early, the larger solution space will contain a lot of inferior cycles, which stochastic solvers (such as QAOA) might pick by accident. Allowing empty trips only later in the algorithm pushes the solver to start by picking cycles entirely composed of full trips, which tend to be better than cycles with (too many) empty trips.

As with QAOA and any sampling algorithm, there is a dependence on the number of shots taken. In this work, we used a constant value of 10,000 shots. The results are pruned to independent sets and then the best result is selected. Obviously, with this approach, larger number of shots leads to better results. Also, for small subset sizes, a large number of shots will lead most sampling algorithms to eventually stumble upon the best solution, so it might erase the difference between completely random guessing and QAOA (whose samples are biased towards the solution).

There is quite a lot of freedom in specifying how to select the subgraphs in each iteration (and even how to assign weights to the graph nodes in the first place). The best choice is very problem-specific and we settled on our choice after some trial and error.

\section{Outlook}\label{sec:outlook}

We have developed a hybrid quantum-classical algorithm for the rolling stock planning problem that combines a classical divide-and-conquer framework with a quantum subproblem solver. The approach decomposes a large-scale instance into tractable subgraphs via a maximum-weight independent set formulation, enabling the use of QAOA on near-term quantum computers. This illustrates how near-term quantum resources can be applied to realistic railway optimization problems in a scalable way that can benefit from future improvements in quantum hardware and algorithms.

A key bottleneck encountered in our study is the generation of feasible train cycles. While exhaustive enumeration is manageable for the problem instance considered here, it may become prohibitive for larger timetables. Future work could explore formulations that avoid explicit cycle enumeration, for example by dynamically generating cycles during the optimization process using techniques inspired by column-generation or branch-and-price methods.

Another open question concerns how improvements in the MWIS solver used on each subgraph translate to improvements in the overall solution quality. While our results show a clear dependence on subgraph size, the impact of stronger subgraph solvers remains less clear. In particular, it would be informative to reproduce Fig.~\ref{fig:subset_size_trend} using QAOA-based solvers. The required subgraph sizes are currently beyond the reach of statevector simulators and present-day quantum hardware, but this may become feasible as quantum processors improve.

Finally, the strategy used to select subgraphs offers considerable flexibility. In this work we observed that selecting cycles with many full trips leads to improved performance. This selection criterion intentionally differs from the MWIS objective used to evaluate solutions, allowing the algorithm to explore a different region of the solution space during subgraph construction. However, the optimal selection strategy is likely problem-dependent and remains an open question. Exploring alternative heuristics or adaptive selection strategies could further improve the performance of the hybrid framework.

\section*{Acknowledgments}
The authors thank Matthew Kiser, Francesco Benfenati and Jernej Rudi Fin\v zgar for proofreading the manuscript and providing helpful editorial suggestions.

The authors thank Manfred Rieck and Ronald Bieber from Deutsche Bahn / DB Systel for providing us with management support, delivery capacity and real-world data (list of trips and information about the network).

\clearpage
\bibliography{bibliography}

\clearpage
\appendix 

\section{Number of Cycles} \label{sec:n_cycles}

Here we consider how many cycles there are in our problem as a function of the number of trips and the time window. In particular, we examine the asymptotic scaling.

\subsection{Upper Bound} \label{sec:upper_bound}

We first establish a rigorous but loose upper bound.

Importantly, we have a fixed time window $T=48$ hours and a fixed shortest duration of a trip $t = 1.1$ hours. This means that a cycle can contain at most $\frac{T}{t} \approx 43$ trips. There are $n=190$ trips. Ignoring any causality or locality constraints, this means that there can be at most $n^{\frac{T}{t}}$ different cycles. While astronomically large, this is \emph{polynomial} in $n$, since neither $T$ nor $t$ depend on it.

\subsection{More Realistic Estimation}

More realistically, the number of cycles is much lower. The $n$ trips are distributed among $m=5$ stations, so on average each station only has $n/m = 38$ outgoing trips throughout the 48-hour period. The average trip is closer to $t^\prime \approx 3$ hours in length and the trains only operate roughly $T^\prime \approx 30$ hours out of the 48-hour period. For simplicity we assume that the $n/m = 38$ trips are evenly-distributed at each station (i.e., one trip each approx. $\Delta = \frac{T^\prime}{n/m} =0.8$ hours).

If we index the 38 trips (regardless of station of origin) by integers 1-38, then a cycle would be a sequence of these integers, such that consecutive integers differ by at least $d=\lceil{\frac{t^\prime}{\Delta}}\rceil= 4$.

To count the number of cycles containing $k$ trips, we use a standard combinatorial formula:
$$N_k = \binom{n/m - (k-1)(d-1)}{k}$$
The total number of all cycles is then:
$$
\sum_{k=1}^{\infty} N_k 
= \sum_{k=1}^{\lfloor \frac{n/m +d -1}{d}\rfloor} \binom{n/m - (k-1)(d-1)}{k} $$

Here we fix the upper bound of the sum to the largest $k$ for which the binomial coefficients will be nonzero.

Plugging in the numbers in our use case gives us 299~915, not too far from the 98~557 value we get by actually enumerating all possible cycles (properly accounting for all constraints and the actual data we work with).

\subsubsection{Leading Term in $n$}
Now we focus on the scaling with $n$, so we need to isolate any dependency on it. First, we expand the variable $d \approx \frac{nt^\prime}{mT^\prime}$ (ignoring rounding up/down). This allows us to simplify the upper limit of the sum to
$$
\Big\lfloor \frac{n/m +d -1}{d}\Big\rfloor \approx \frac{n/m +\frac{nt^\prime}{mT^\prime} -1}{\frac{nt^\prime}{mT^\prime}} \approx \frac{T^\prime}{t^\prime}+1
$$

Similarly, the upper term in the binomial coefficient is simplified as
$$
\frac{n}{m} - (k-1)(d-1) = \frac{n}{m} - (k-1) \bigg(\frac{nt^\prime}{mT^\prime}-1\bigg) \approx \frac{n}{m}\Big[ 1-(k-1)\frac{t^\prime}{T^\prime} \Big]
$$

This allows us to upper bound the binomial coefficient as

$$
\binom{\frac{n}{m}\big[ 1-(k-1)\frac{t^\prime}{T^\prime} \big]}{k} \leq \frac{\frac{n^k}{m^k}\Big[ 1-(k-1)\frac{t^\prime}{T^\prime} \Big]^k}{k!}
$$
Now all the terms not dependent on $n$ can be ignored, so we end up with scaling $\mathcal{O}(n^k)$. Given the range for $k$, this can be at most $\mathcal{O}(n^\frac{T^\prime}{t^\prime})$, similar to the result obtained in subsection  \ref{sec:upper_bound}, although now the exponent is much smaller and there are additional small constant factors hidden in the $\mathcal{O}$-notation.

\end{document}